\documentclass[]{svmult}

\usepackage{mathptmx}       
\usepackage{makeidx}         
\usepackage{graphicx}        
\usepackage[bottom]{footmisc}

\makeindex             

\def\ts     {\thinspace}
\def\kms    {\ifmmode{{\rm \ts km\ts s}^{-1}}\else{\ts km\ts s$^{-1}$}\fi}
\def\msol   {\ifmmode{{\rm M}_{\odot}}\else{M$_{\odot}$}\fi}
\def\lsol   {\ifmmode{{\rm L}_{\odot}}\else{L$_{\odot}$}\fi}
\def\zsol   {\ifmmode{{\rm Z}_{\odot}}\else{Z$_{\odot}$}\fi}
\def\etal   {{\rm et\ts al. }}

\begin{document}

\title*{Cosmic evolution of gas content and accretion}
\author{Fran\c{c}oise Combes}
\institute{Observatoire de Paris, LERMA, CNRS, 61 Av. de l'Observatoire, F-75014, Paris, France \email{francoise.combes@obspm.fr}}
%
%
\maketitle
\vspace{-2cm}

\abstract{In the present universe, the gas is a minor component
of giant galaxies, and its dominant phase is atomic (HI). During galaxy 
evolution in cosmic times, models predict that gas fractions were much higher 
in galaxies, and gas phases could be more balanced between molecular (H$_2$)
and atomic (HI). This gaseous evolution is certainly a key factor to
explain the cosmic evolution of the star formation rate density. 
Star formation efficiency might also vary with redshift, and the relative
importance of these factors is not yet well known. Our current knowledge
of cosmic evolution of gas from molecular observations at high-z is
reviewed and confronted to simulations.}

\section{Introduction}
\label{sec:1}

The abundance of gas observable in galaxies today can be expressed
with dimensionless numbers, normalised with the critical density of the 
universe. While stars in galaxies account for $\Omega_*$ = 3 10$^{-3}$
(e.g. Fukugita \etal 1998), the HI gas contributes by  
$\Omega_{HI} \sim$ 3.5 10$^{-4}$ (Zwaan \etal 2005),
and the molecular gas by $\Omega_{H_2} \sim$ 1.2 10$^{-4}$ (Young \etal 1995,
Sauty \etal 2003, Keres \etal 2003, Saintonge \etal 2011).

 Theoretical considerations and semi-analytical models predict that the molecular-to-atomic
 gas ratio decreases regularly with cosmic time in galaxies 
(Obreschkow \& Rawlings, 2009, Obreschkow \etal 2009).
 The phase transition to molecular hydrogen can be formulated in terms of
pressure (Blitz \& Rosolowsky 2006), and the surface density and 
consequently the pressure is higher in high-z galaxies. The modelisation
leads to a dependency of H$_2$/HI varying as (1+z)$^{1.6}$.
This is essentially due to the expectation that the size of galaxies
is growing as (1+z)$^{-1}$ with cosmic time.

  The evolution with z of $\Omega_{HI}$ in galaxies is not yet known from
emission, but can be derived from the damped Lyman-$\alpha$ absorption in front
of quasars, since these systems are thought to correspond to galaxies.
 Albeit with large error bars, the abundance of HI appears about constant
from z=4 to z=0 (Zwaan \etal 2005). It is however expected to experience
a strongly varying phase at higher z, when cold gas settles in galaxies,
through accretion and cooling, mergers, etc. At these early epochs, 
molecules might have difficulties to form, since metals and dust are 
building up slowly, but the exact processes are not yet well known.

What is better known is the cosmic evolution of star formation rate density,
from UV to far-infrared light, and its decrease by a factor 20 since z=2
(e.g. Hopkins \& Beacom 2006, Bouwens \etal 2011). How does this SFR evolution
relate to the cosmic cold gas evolution? Is it linked to HI or H$_2$ density,
or/and to the star formation efficiency (SFE)?

\section{High-z molecular observations}
\label{sec:2}

Since about 20 years now, molecular gas is observed in high redshift galaxies.
Due to the lack of sensitivity, mostly lensed galaxies were first discovered
(cf the review  by Solomon \& vanden Bout 2005). More and more "normal" objects,
on the main sequence of star forming galaxies are observed now, and this will increase
considerably with ALMA. The detection of CO lines at high redshift is made easier 
by the existence of the rotational ladder, where the flux of the higher transitions
can be much higher than the fundamental line. This is not the sase of the HI gas and
the only 21cm line, which will have to wait SKA to be detectable at high redshift.

\subsection{Starbursts and ULIRGs}
\label{ULIRG}

Until very recently, only very luminous galaxies in the far-infrared (LIRGs and ULIRGs)
were detected in the CO lines at high redshift, due to limited sensitivity. In
the local universe, it is now well established that ULIRGs are starbursts triggered
by galaxy interactions and mergers (e.g. Solomon \etal 1997). At high redshift,
the global star formation rate is increasing rapidly, and even ULIRG are
not all starbursts. It is thought now that the starburst mode is likely to represent
only 10\% of the stars formed at z=2, the cosmic peak of the star formation activity
(Rodighiereo \etal 2011).

Already Greve \etal (2005) showed that the SFE (defined by the ratio of FIR luminosity, 
taken as an indicator of SFR, to the CO luminosity, indicator of the gas mass) was 
increasing significantly with redshift, reaching maxima around z=2 for submillimeter galaxies
(SMG) with an SFE up to 2 orders of magnitude higher than for local LIRGs. 
The gas consumption time-scale, being the inverse of the  SFE, could then fall to
20 Myrs, instead of the average 2 billion yrs locally.

\begin{figure}[ht]
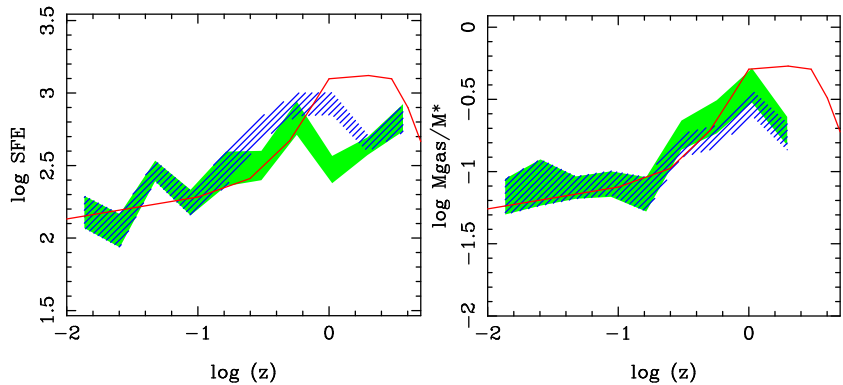

\includegraphics[width=5.5cm]{SFE-fill.ps}
\includegraphics[width=5.5cm]{GASF-fill.ps}
\caption{{\bf Left}: Evolution of the star formation efficiency (SFE) with redshift.
{\bf Right}: Cosmic evolution of the gas to stellar mass ratio, for the LIRG and ULIRG
compilation of Combes \etal (2013). The green area corresponds to the CO-detected points,
and the hatched area also includes the 3$\sigma$ upper limits. The width of the shaded
regions correspond to the statistical scatter in N$^{-1/2}$. The red curve is indicative 
of the logarithmic variations of the cosmic star formation rate density (Hopkins \& Beacom 2006).}
\label{fig:galevol}  
\end{figure}

The redshift range between z=0.2 and z=1 is very important for the cosmic gas evolution,
since it is the period when the cosmic star formation density drops by a factor 10,
and it corresponds to 40\% of the universe age.
Unfortunately, this domain was not easily observed because of atmospheric lines, 
and the need of sensitive 2mm instruments.
A sample of 69 ULIRG was observed in different CO lines with the IRAM-30m
precisely in this redshift range (Combes \etal 2011, 2013).
 From the galaxies where the gas excitation is known, and from the dust masses derived
from the continuum emission, the adoption of the ULIRG CO-to-H$_2$ conversion factor
is justified (e.g. Solomon \etal 1997). This ratio is 5.7 times smaller than the standard
ratio adopted for Milky Way-like galaxies. The average molecular mass is however 
1.45 10$^{10}$ M$_\odot$, an order of magnitude higher than in the Milky Way.

Compiling this sample with other LIRGs and ULIRGs, both the molecular 
gas to stellar mass ratio and the SFE significantly 
increase with redshift, by factors of $\sim$ 3 from z = 0 to 1, 
as shown in Figure \ref{fig:galevol}, suggesting that
both factors play an important role and complement each other in cosmic star formation evolution.

\begin{figure}[ht]
\centerline{
\includegraphics[width=9cm]{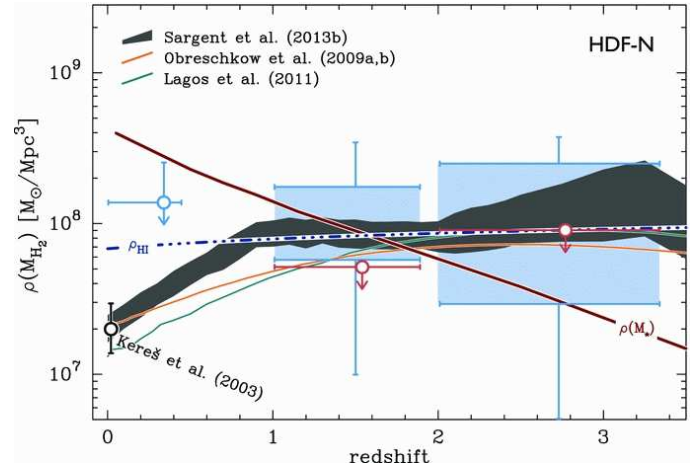}}
\caption{Evolution of the cosmic H$_2$ mass density versus redshift,
comparing observational limits obtained from
blind detections in the Hubble Deep Field North by Decarli \etal (2014) 
shown in blue-shaded areas, to predictions 
from semi-analytical cosmological models (Obreschkow \etal 2009; 
Lagos \etal 2011) and empirical predictions by Sargent \etal 2014
(grey-shaded areas). The red upper limit corresponds only
to galaxies selected via optical spectroscopic redshifts. 
The evolution of the atomic gas mass 
density ($\rho_{HI}$) and of the stellar mass density ($\rho$(M$_*$)) are 
also plotted (from Walter \etal 2014).}
\label{fig:h2z-walter}
\end{figure}

\subsection{Main sequence galaxies}
\label{MS}

Not all star forming galaxies at z=1-2 have a high SFE. Some galaxies,
selected only from their optical colors, were detected
in the CO lines with surprising high CO luminosities (Daddi \etal 2008).
These galaxies, although still in the ULIRGs range, have a low gas excitation
(Dannerbauer \etal 2009), and are relatively extended. They are interpreted as
disk-like galaxies with steady star formation rate, while the most excited
ULIRGs are nuclear starbursts. It is possible that the Milky Way-like 
conversion ratio applies for these objects, which will further lower
their SFE. However, the adoption of a bimodal conversion ratio
leads to an artificial bimodal star formation regime, separating
the starbursts from the more quiescent disks with a gap of an order
of magnitude in gas consumption time-scales. In reality, there must
exist a continuous conversion ratio, according to gas density, temperature,
and other factors like metallicity, and the SF regimes are certainly continuous too.

A continuity of galaxy properties between the two modes of star formation, main sequence
and starburst, is developed further by Daddi \etal (2013) and Sargent \etal (2014). Although
starbursts have larger SFE, it is not easy to know whether the cause is a lack of gas
(may be the consequence of a short boost of star formation), or an excess of young stars.
 If the starburst is triggered by a merger, numerical simulations show that
gas is driven inwards by gravity torques from the outer reservoir, and more gas 
is then observable (e.g. Di Matteo \etal 2007, Montuori \etal 2010). An excess of fresh
gas in star forming galaxies
is also supported by the fundamental mass-metallicity relation, which
precisely depends on SFR (Manucci \etal 2010). Starbursts have also a larger
molecular gas to stellar mass ratio, so their elevated SFR is both due to 
a larger gas content and a larger SFE. The latter could be due to the larger 
central concentration of the gas, and this will be clarified through resolved
SFR/gas density studies.

The PHIBSS large program on the IRAM interferometer (Tacconi \etal 2010, 2013,
see also this conference)
has targeted a sample of massive star forming galaxies, likely to be on the 
main sequence as defined in the stellar mass-SFR diagram (e.g. Wuyts \etal 2011).
From the 52 CO-detected objects at z=1-3, the gas mass fraction is found to 
increase with z, up to 50\%, and decrease with mass. Most of the objects look
like disks with regular rotation, and are more steady star forming disks than starbursts,
without any interaction or merger. Since the molecular gas depletion time-scale
is typically 0.7 Gyr and varies as (1+z)$^{-1}$, the star formation must be fueled by gas 
accretion episodes, which are frequent at high and moderate
redshift (e.g. Combes 2014). The resolved Kennicutt-Schmidt
relation obtained in a few objects is compatible with a linear relation, and a
depletion time-scale lower at high-z (Freundlich \etal 2013, Genzel \etal 2013).

In all these massive star forming galaxies, atomic gas cannot be dominating the cold gas,
since the sum of the molecular and stellar masses are so close to the dynamical mass.
Unless the CO-to-H$_2$ conversion factor is largely in error, the H$_2$/HI
ratio has indeed increased with z, as predicted by models. Another recent study supports
this conclusion: Decarli \etal (2014) have carried out a blind molecular line survey in the
Hubble Deep Field North, scanning the whole 3mm band with the IRAM interferometer.
 Their blind detection of 17 CO lines, together with the upper limit
obtained by stacking the observations towards spectroscopically identified
objects, constrain the CO luminosity functions at the corresponding redshifts.  They
deduce that optical/MIR bright galaxies contribute less than 50\% to the star formation 
rate density at 1 $<$ z $<$ 3.
 Their derived evolution of the H$_2$ mass density is compared to models in
Figure \ref{fig:h2z-walter}.

\begin{figure}[ht]
\centerline{
\includegraphics[width=8cm]{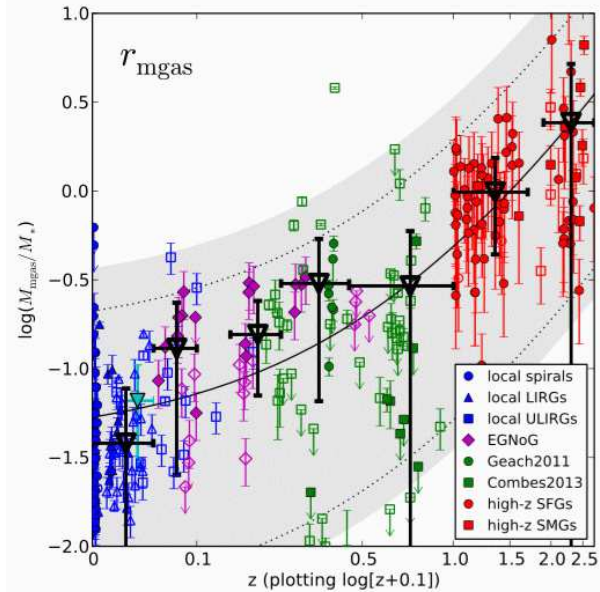}}
\caption{Evolution of the molecular gas to stellar mass ratio (r$_{mgas}$) versus z, from
the compilation of Bauermeister \etal (2013). Symbols are filled for main sequence
galaxies, and empty for starbursts. The 7 bold black triangles are the average
for the different redshift bins. The shaded grey zone indicates the expected
region for normal galaxies, with the solid curve being the average.}
\label{fig:EGNOG}
\end{figure}

A recent 870$\mu$m continuum survey with ALMA of SMG in the Extended Chandra Deep Field South 
(Swinbank \etal 2014) has discovered that the well detected sources (S$_{870}>$ 2mJy) are
in average ULIRGs with SFR=300 M$_\odot$/yr. The extrapolation of the counts down to
 S$_{870}>$ 1mJy show that these sources contribute to 20\% of the 
cosmic star formation density over z=1-4. Deriving H$_2$ masses from dust masses,
the average SFE is found rather high, with depletion time-scale of 130 Myr.
This is to be compared to the compilation by Bauermeister \etal (2013),
who observed normal star forming galaxies in the redshift range z=0.05-0.5. 
They find a depletion time-scale for normal galaxies of 760 Myr,
and for starbursts 60 Myr. Their derived molecular gas to stellar mass ratio
is plotted in Figure \ref{fig:EGNOG}, and is compatible with the model-expected behavior.

\section{Models and simulations}
\label{sec:3}

Semi-analytical models (SAM) have been run, within the standard $\Lambda$CDM model, to
compute the cosmic evolution of the cold gas content in galaxies. Lagos \etal
(2011) show that the best recipe to control the phase transition from atomic to molecular is
the pressure model (Blitz \& Rosolowsky 2006), rather than the theory-based model from
Krumholz \etal (2009) taking into account UV-dissociation of molecules and their
reformation on grains. In their best fit model, the H$_2$/HI ratio 
rises above one at high redshift, as in Obreschkow \etal (2009).
Fu \etal (2010, 2012) claim that the Krumholz \etal (2009) recipe is better, but on a limited 
mass range. Their best fit requires that the depletion time-scale
remains 1-2 Gyr at high redshift.
 
Using a simple phenomenological model, 
Feldmann (2013) claims that the relation between SFR and H$_2$ content is likely
to be linear at all redshift. This assumption provides the best fit to the data, 
i.e. the cosmic star formation history, the evolution of the
mass-metallicity relation, and the gas-to-stellar mass ratio in galaxies.
 This means that the variation of SFE with redshift might be too little to be sensitive.
  Models where the SFR relation is non-linear with gas density produce too much 
stars and metals early on to be compatible with the observations.
 To obtain the right star formation histories, gas accretion must be limited to 
a halo mass range between a critical minimum mass M$_c$(z), below which
 photoionisation limits the baryon fraction, and an upper limit 
 M$_{shock} \sim$2 10$^{12}$ M$_\odot$, above which the gas is heated by shocks
in entering the galaxy (Birnboim \& Dekel, 2003).
  At early epochs, for redshifts higher than 2, the gas accretion time-scale is very short, 
and the SFR not enough to consume the accreted gas, which accumulates in galaxies.
 After z=2, the SFR has increased to its maximum; within the halo mass range between 
M$_c$ and  M$_{shock}$, the depletion time-scale is comparable to the accretion time-scale, and
 the SFR is limited by gas accretion (Bouch\'e \etal 2010).
 In this global model, an equilibrium settles between gas inflow and outflow,
and star formation rate, equalling depletion time to accretion time. Stellar
masses, metallicity, and cosmic gas evolution are moderated by this equilibrium.

The relation between SFR and stellar mass on the main sequence has
been examined in detail from 25 studies in the literature, and the corresponding slope
is a decreasing function of cosmic time (Speagle \etal 2014). The star formation histories 
derived from these are delayed-$\tau$ models, where the SFR is first increasing 
linearly with time in the first half of the universe age, and then decreasing exponentially.

With a SAM approach Popping \etal (2012, 2014) also tested several recipes for
the molecular gas and star formation evolution; either pressure-based, or metallicity-based models
represent rather well observations, with some variations for low mass galaxies.
To compare with high-z observations, they deduce their gas content from their SFR,
through inversion of the Kennicutt-Schmidt (KS) relation, but in this case the best fit
is for a density-dependent SFE. Also, the CO-to-H$_2$ conversion factor should
be continuous, as a function of the galaxy physical properties.

That the SFE should depend on gas surface density (non-linear KS relation) is certainly a solution
to explain why SFE varies with redshift.
Galaxies were more compact at high z (Newman \etal 2012, Morishita \etal 2014),
so not only their surface density was higher for a given gas content, but also their
dynamical time was shorter, which favors the dynamical triggers.
Another feature is the volumic density dependency, which could play a role,
even for a linear KS relation. It has already been observed that SFE is declining with
radius in galaxy disks at z=0, possibly due to gas disk flaring
(Bigiel \etal 2010, Dessauges-Zavadsky \etal 2014).

\bigskip
There are still large uncertainties on key factors to determine the cosmic
evolution of cold gas content in galaxies: not only the SFR laws as a function of
density, the phase transition between atomic and molecular gas,
but also the star formation efficiency, regulated by feedback mechanisms
due to supernovae or AGN, the quenching due to environment, slowing down
the gas accretion. We are just at the beginning of the ALMA era,
and our knowledge on these physical processes will progress exponentially.

\begin{acknowledgement}
My great thanks to the organisers, David Block, Ken Freeman and Bruce Elmegreen
 for this wonderful meeting with wide scientific interests.
The European Research Council is gratefully  acknowledged
for the Advanced Grant Program Num 267399-Momentum.
\end{acknowledgement}

\parindent=0pt
{\bf References}
\parindent=0pt

{\small
 
Bauermeister A., Blitz L., Bolatto A., \etal  2013, ApJ, 768, 132
\\ Bigiel, F., Leroy, A., Walter, F., \etal 2010,  AJ, 140,  1194
\\ Birnboim Y., Dekel A.: 2003 MNRAS 345, 349
\\ Blitz L., Rosolowsky E.: 2006, ApJ 650, 933
\\ Bouch\'e N., Dekel A., Genzel R. \etal 2010, ApJ 718, 1001
\\ Bouwens, R. J., Illingworth, G. D., Oesch, P. A. \etal : 2011 ApJ 737, 90
\\ Combes F., Garc{\'{\i}}a-Burillo S., Braine J. \etal 2011, A\&A, 528, A124
\\ Combes F., Garc{\'{\i}}a-Burillo S., Braine J. \etal 2013, A\&A, 550, A41 
\\ Combes F., 2014, Arkansas conf. , arXiv:1309.1603 
\\ Daddi E., Dannerbauer, H., Elbaz, D. \etal 2008, ApJ 673, L21
\\ Daddi E., Sargent M.~T., B{\'e}thermin M., Magdis G., 2013, IAUS, 295, 64 
\\ Dannerbauer, H., Daddi, E., Riechers, D. A. \etal 2009, ApJ 698, L178
\\ Decarli, R., Walter, F., Carilli, C. \etal 2014 ApJ 782, 78
\\ Dessauges-Zavadsky M., Verdugo C., Combes F., Pfenniger D.: 2014, A\&A in press
\\ Di Matteo, P., Combes, F., Melchior A-L., Semelin, B.: 2007, A\&A 468, 61
\\ Feldmann R. 2013, MNRAS 433, 1910
\\ Freundlich J., Combes F., Tacconi L. \etal 2013, A\&A 553, A130
\\ Fu, J., Guo, Q., Kauffmann, G., Krumholz, M. R. 2010,  MNRAS 409, 515
\\ Fu, J., Kauffmann, G., Li, C., Guo, Q. 2012, MNRAS 424, 2701
\\ Fukugita M., Hogan C. J., Peebles P. J. E., 1998, ApJ, 503, 518
\\ Genzel, R., Tacconi, L. J., Kurk J. \etal 2013,  ApJ 773, 68
\\ Greve. T. R., Bertoldi, F., Smail, I. \etal  2005, MNRAS, 359, 1165
\\ Hopkins A. M., Beacom J. F.: 2006, ApJ 651, 142
\\ Keres, D., Yun, M. S., Young, J. S. 2003, ApJ, 582, 659
\\ Krumholz M. R., McKee C. F., Tumlinson J.: 2009 ApJ 699, 850
\\ Lagos C. d P., Baugh C. M., Lacey C. G. \etal 2011, MNRAS 418, 1649
\\ Mannucci, F., Cresci, G., Maiolino, R. \etal 2010, MNRAS 408, 2115
\\ Montuori M., Di Matteo, P., Lehnert, M. D., Combes, F., Semelin, B.: 2010, A\&A 518, A56
\\ Morishita T., Ichikawa, T., Kajisawa, M.: 2014, ApJ 785, 18
\\ Newman A. B., Ellis, R. S., Bundy, K., Treu, T. : 2012, ApJ 746, 162
\\ Obreschkow D., Croton D., de Lucia G. \etal 2009, ApJ 698, 1467
\\ Obreschkow D., Rawlings S.: 2009, ApJ 696, L129
\\ Popping G., Caputi K.~I., Somerville R.~S., Trager S.~C., 2012, MNRAS, 425, 2386 
\\ Popping G., Somerville R.~S., Trager S.~C., 2014, arXiv:1308.6764 
\\ Rodighiero, G., Daddi, E., Baronchelli, I. \etal 2011, ApJ 739, L40
\\ Saintonge, A., Kauffmann, G., Kramer, C. \etal 2011, MNRAS 415, 32
\\ Sargent, M. T., Daddi, E., Bethermin, M  \etal : 2014, ApJ in press, arXiv1303.4392
\\ Sauty S., Casoli, F., Boselli, A. \etal 2003, A\&A, 411, 381
\\ Solomon P., Downes D., Radford S., Barrett J.: 1997, ApJ 478, 144
\\ Solomon, P. M., Vanden Bout, P. A.: 2005, ARA\&A 43, 677
\\ Speagle J. S., Steinhardt C. L., Capak P. L., Silverman J. D: 2014, ApJ sub arXiv1405.2041
\\ Swinbank A.~M., Simpson J. M., Smail I., \etal 2014, MNRAS, 438,  1267 
\\ Tacconi L. J., Genzel R., Neri R.  \etal 2010 Nature 463, 781
\\ Tacconi L. J., Neri R., Genzel R. \etal 2013, ApJ 768, 74
\\ Walter F., Decarli R., Sargent M. \etal 2014, ApJ,  782, 79
\\ Wuyts, S., F\"orster Schreiber, N. M., van der Wel, A.  \etal 2011, ApJ 742, 96
\\ Young, J. S., Xie, S., Tacconi, L. \etal  1995, ApJS 98, 219
\\ Zwaan, M. A., Meyer, M. J., Staveley-Smith, L., Webster, R. L.: 2005, MNRAS 359, L30
}
\end{document}